\documentclass[useAMS,usenatbib]{mn2e}

\title[C, N and O abundances in red clump stars]
      {C, N and O abundances in red clump stars of the \\ Milky Way}

\author[G. Tautvai\v sien\.{e} et al.]
    { G. Tautvai\v sien\. e,$^{1}$\thanks{E-mail:grazina.tautvaisiene@tfai.vu.lt} B. Edvardsson,$^{2}$ E. Puzeras,$^{1}$
G. Barisevi\v{c}ius$^{1}$ and I. Ilyin$^{3}$\\
       $^{1}$Institute of Theoretical Physics and Astronomy, Vilnius University, Go\v{s}tauto 
12, Vilnius 01108, Lithuania\\
       $^{2}$Department of Physics and Astronomy, Uppsala Astronomical Observatory, Box 515, 751\,20 Uppsala, Sweden\\ 
       $^{3}$Astrophysikalisches Institut Potsdam, An der Sternwarte 16, Potsdam 14482, Germany }

\begin{document}

\date{Accepted 2010 ..... Received 2010 ......; in original form 2010 ......}

\pagerange{\pageref{firstpage}--\pageref{lastpage}} \pubyear{2010}

\maketitle

\label{firstpage}

\begin{abstract}
The {\it Hipparcos} orbiting observatory has revealed a large number of helium-core-
burning "clump" stars in the Galactic field. These low-mass stars exhibit signatures
of extra-mixing processes that require modeling beyond the first dredge-up of standard models. 
The $^{12}{\rm C}/^{13}{\rm C}$ ratio is the most robust diagnostic of deep mixing, because it 
is insensitive to the adopted stellar parameters. 
In this work we present $^{12}{\rm C}/^{13}{\rm C}$ determinations in a sample of 34 Galactic clump 
stars as well as abundances of nitrogen, carbon and oxygen. 
Abundances of carbon were studied using the ${\rm C}_2$ 
Swan (0,1) band head at 5635.5~{\AA}.  The wavelength interval 
7980--8130~{\AA} with strong CN features was analysed in order to determine 
nitrogen abundances and $^{12}{\rm C}/^{13}{\rm C}$  isotope ratios. 
The oxygen abundances were determined from the [O\,{\sc i}] line at 6300~{\AA}. 
Compared with the Sun and dwarf 
stars of the Galactic disk, mean abundances in the investigated clump stars 
suggest that carbon is depleted by about 0.2~dex, nitrogen is enhanced by 0.2~dex and oxygen is 
close to abundances in dwarfs.  
Comparisons to evolutionary models show that the stars fall into two groups: the one is of first ascent 
giants with carbon isotope ratios altered according to the first dredge-up prediction, and the other 
one is of helium-core-burning stars with carbon isotope ratios altered by extra mixing. 
The stars investigated fall to these groups in approximately equal numbers. 

\end{abstract}


\section{Introduction}

During the last decades, an increasing amount of work has been done in studying the chemical 
composition of red clump stars of the Galaxy (e.g. McWilliam 1990; Tautvai\v{s}ien\.{e} et al.\ 2003; 
Mishenina et al.\ 2006; Liu et al.\ 2007; Luck \& Heiter 2007; Tautvai\v{s}ien\.{e} \& Puzeras 2009; 
Puzeras et al.\ 2010). 
From the {\it Hipparcos} catalogue (Perryman et al.\ 1997) containing about 600 clump stars with 
parallax error lower than 10\% and representing a complete sample of clump stars to a distance 
of about 125~pc, almost a half of stars are already investigated by means of high resolution 
spectroscopy. 

Among the fundamental questions to which investigations of clump stars should help 
to find an answer is a mechanism of transport of processed material to the stellar surface in 
low mass stars. Post-main sequence stars with masses below $2-2.5~M_{\odot}$ exhibit signatures 
of material mixing that require challenging modelling beyond the standard stellar theory 
(reviews by Charbonnel 2006; Chanam\'{e} et al.\ 2005 and references therein). Also it is interesting 
to find out how many stars in the Galactic clump belong to the first ascent giants and to 
He-core-burning stars. 

Carbon and nitrogen abundances are among most useful quantitative indicators of 
mixing processes in evolved stars. Because of the first dredge-up abundances of 
$^{12}{\rm C}$ decrease while abundances of $^{13}{\rm C}$ and $^{14}{\rm N}$ increase (Iben 1965). 
Depending on stellar mass, metallicity and evolutionary state, these alterations are growing 
(c.f. Boothroyd \& Sackmann 1999; Charbonnel \& Zahn 2007; and many other studies).
 
Three large studies of $^{12}{\rm C}$, N and O abundances in clump stars have been done recently. 
Abundances of C, N and O in 177 
clump giants of the Galactic disk were determined by Mishenina et al.\ (2006) on 
a basis of spectra ($R$=42\,000) obtained on the 1.93-m telescope of the Haute-Provence Observatoire 
(France).   

A sample of 63 red clump stars, mainly located in the southern hemisphere, was investigated by 
Liu et al.\ (2007). Abundances of oxygen were investigated on 
a basis of spectra ($R$=48\,000) obtained on the 1.52-m telescope of the ESO (La Silla, Chile).  

A spectroscopic analysis of C, N and O ($R$=60\,000) was done for a sample of nearby giants, with red 
clump stars among them by Luck \& Heiter (2007). We selected a subsample of 138 red clump giants based on 
the luminosity and effective temperature diagram in the Fig.~20 of this paper. All the stars located 
in the box limited by luminosities log($L/L_{\odot}$) from 1.5 to 1.8 and effective temperatures 
from 4700~K to 5200~K were includes in the subsample.         

A comprehensive study of $^{13}{\rm C}$ abundances in Galactic clump stars was not done yet. 
The $^{12}{\rm C/}^{13}{\rm C}$ ratio is the most robust diagnostic of deep mixing, because it 
is very sensitive to mixing processes and is almost insensitive to the adopted stellar parameters. 

In this paper we report $^{12}$C, $^{13}$C, N and O abundances 
in the 34 clump stars of the Galactic field obtained from the high-resolution spectra. 
The results are discussed in detail together with results of other studies of the clump stars. 
The preliminary results of this study were published by Tautvai\v{s}ien\.{e} et al.\ (2003, 2007, 2010) 
and Tautvai\v{s}ien\.{e} \& Puzeras (2009).

\section{Observational data}

The spectra of 26 stars were obtained at the Nordic Optical Telescope (NOT, La Palma) 
with the SOFIN \'{e}chelle spectrograph (Tuominen et al. 1999).  
The 2nd optical camera ($R\approx 80\,000$) was used to observe simultaneously 13 spectral orders, 
each of $40-60$~{\AA} in length, located from 5650~{\AA} to 8130~{\AA}. 
Reduction of the CCD images, obtained with SOFIN, was done using the {\em 4A} software package 
(Ilyin 2000). Procedures of bias subtraction, spike elimination, flat field 
correction, scattered light subtraction, extraction of spectral orders were 
used for image processing. A Th-Ar comparison spectrum was used for the 
wavelength calibration. The continuum was defined from a number of narrow 
spectral regions, selected to be free of lines.

This sample of stars was supplemented by spectroscopic observations ($R\approx 37\,000$) of 8 red clump stars 
obtained on the 2.16~m telescope of the Beijing Astronomical Observatory (China) taken from the literature 
(Zhao et al.\ 2001). There are more spectra of clump stars presented in this literature source, however not 
all of them have regions of ${\rm C}_2$ Swan (0,1) band head at 5630.5~{\AA} observed, or good quality 
$^{13}{\rm C}^{14}{\rm N}$ bands at 8004~\AA. In Fig.~1, we show examples of spectra observed on the 
Nordic Optical Telescope and the telescope of Beijing Astronomical Observatory.  
 
\section{Method of analysis and physical data}

The spectra were analysed using a differential model atmosphere technique. 
A program {\sc bsyn}, developed at the Uppsala Astronomical 
Observatory, was used to carry out the calculations of synthetic spectra. A set of plane parallel, 
line-blanketed, constant-flux LTE model atmospheres 
was computed with an updated version of the {\sc marcs} code (Gustafsson et al.\ 2008).
Calibrations to the solar spectrum (Kurucz et al.\ 1984) was done for all the spectral regions investigated. 
For this purpose  we used the 
solar model atmosphere from the set calculated in Uppsala with a microturbulent 
velocity of 0.8~$\rm {km~s}^{-1}$, as derived from 
Fe~{\sc i} lines, and the solar abundances log$A_{\rm C} = 8.52$,  
log$A_{\rm N} = 7.92$, log~$A_{\rm O}=8.83$, log~$A_{\rm Fe}=7.50$, ${\rm C/N} = 3.98$, 
$^{12}{\rm C/}^{13}{\rm C} = 89$, etc.\ (Grevesse \& Sauval 2000). 

\input epsf
\begin{figure}
\epsfxsize=\hsize 
\epsfbox[15 10 170 130]{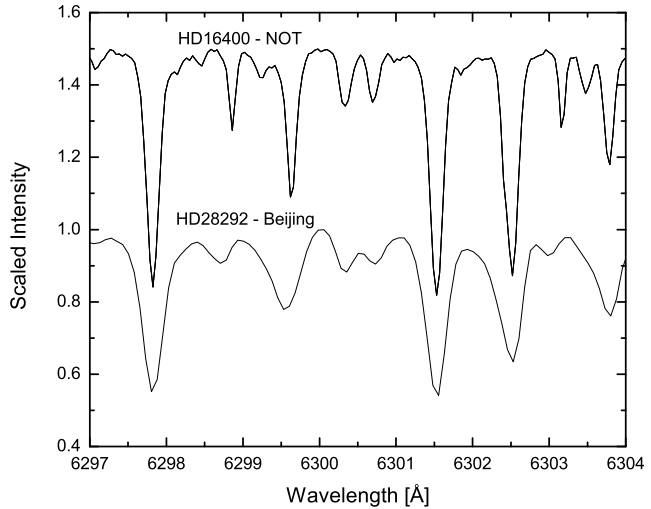} 
\caption{Stellar spectrum examples in the region of [O\,{\sc i}] line $\lambda 6300.3$~\AA\ 
observed on the Nordic Optical Telescope (NOT) ($R\approx 80\,000$) and at the Beijing Astronomical 
Observatory ($R\approx 37\,000$).} 
\label{fig1}
\end{figure}

For  ${\rm C}_2$ determination in stars we used 5632 -- 5636~{\AA} interval to compare 
with observations of ${\rm C}_2$ Swan (0 ,1) band head at 5635.5~{\AA}. The same 
molecular data of ${\rm C}_2$ as used by Gonzalez et al.\ (1998) were adopted for the analysis.
This feature was used in several of our 
previous studies of giants (Tautvai\v{s}ien\.{e} et al.\ 2000; 2001; 2005).
 
The interval 7980 -- 8130~{\AA} contains strong $^{12}{\rm C}^{14}{\rm N}$ and $^{13}{\rm C}^{14}{\rm N}$ 
features, so it was used for nitrogen abundance and $^{12}{\rm C}/^{13}{\rm C}$ ratio analysis. 
The well known $^{13}{\rm CN}$ line at 8004.7~\AA\ was analysed in order to determine $^{12}{\rm C}/^{13}{\rm C}$ ratios. 
The molecular data for this CN band were provided by Bertrand Plez (University of Montpellier II). 
All $gf$ values of CN were increased by +0.03~dex to fit the model spectrum of solar atlas of Kurucz et. al.\ (1984).

We derived oxygen abundance from synthesis of the forbidden [O\,{\sc i}] line at 6300~{\AA}. 
The $gf$ values for $^{58}{\rm Ni}$ and $^{60}{\rm Ni}$ isotopic line components, which blend the 
oxygen line, were taken from Johansson et al.\  (2003) and [O\,{\sc i}]  
log~$gf = -9.917$ value, as calibrated to the solar spectrum (Kurucz et al.\ 1984).

The atomic oscillator strengths for stronger lines of iron and other elements were taken from 
Gurtovenko \& Kostik (1989). The Vienna Atomic Line Data Base (VALD, Piskunov et al.\ 1995) was extensively 
used in preparing the input data for the calculations. 
In addition to thermal and microturbulent Doppler broadening of lines, atomic 
line broadening by radiation damping and van der Waals damping were considered 
in the calculation of abundances. Radiation damping parameters of 
lines were taken from the VALD database. 
In most cases the hydrogen pressure damping of metal lines was treated using 
the modern quantum mechanical calculations by Anstee \& O'Mara (1995), 
Barklem \& O'Mara (1997) and Barklem et al.\ (1998). 
When using the Uns\"{o}ld (1955) approximation, correction factors to the classical 
van der Waals damping approximation by widths 
$(\Gamma_6)$ were taken from Simmons \& Blackwell (1982). For all other species a correction factor 
of 2.5 was applied to the classical $\Gamma_6$ $(\Delta {\rm log}C_{6}=+1.0$), 
following M\"{a}ckle et al.\ (1975). 

Stellar rotation was taken into account when needed. The values of $v {\rm sin} i$ have 
been taken from Hekker \& Mel\'{e}ndez (2007), De Medeiros et al.\ (2002) and Glebocki \& Stawikowski (2000). 

Effective temperature, gravity, [Fe/H]  and microturbulent velocity values of the stars have 
been taken from Puzeras et al.\ (2010) where 
these values were derived using traditional spectroscopic criteria. 

Determinations of stellar masses were performed using effective temperatures obtained by Puzeras et al., luminosities 
and Girardi et al.\ (2000) isochrones. The luminosities were calculated from Hipparcos 
parallaxes (van Leeuwen 2007), $V$ magnitudes (SIMBAD database), bolometric corrections calculated according to 
Alonso et al.\ (1999) and interstellar reddening corrections calculated using Hakkila et al.\ (1994) software.
    
\input epsf
\begin{figure}
\epsfxsize=\hsize 
\epsfbox[15 10 170 135]{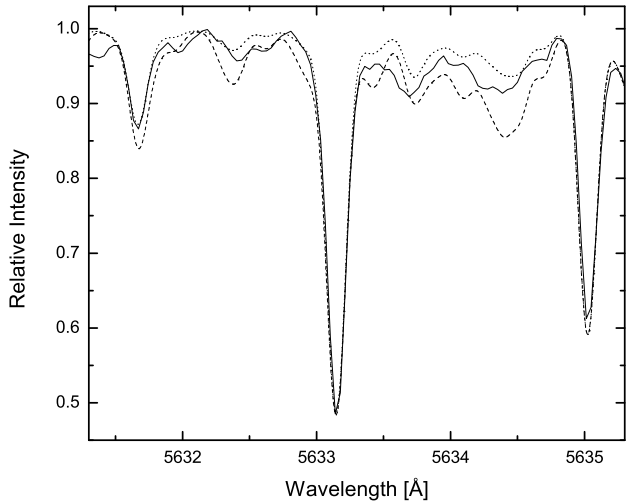} 
\caption{Synthetic and observed spectra for the C$_2$ region near $\lambda 5635$~\AA\ of HD\,8763. 
The solid line shows the observed spectrum and the dotted and dashed lines show the synthetic 
spectra generated with ${\rm [C/Fe]}=0$ and $-0.2$.} 
\label{fig2}
\end{figure}

\subsection{Estimation of uncertainties}

\input epsf
\begin{figure}
\epsfxsize=\hsize 
\epsfbox[25 200 565 590]{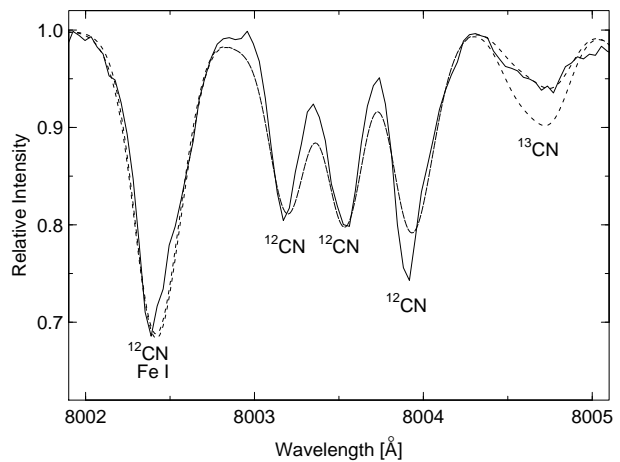} 
\caption{Stellar spectrum synthesis example around CN lines in HD\,2910. The solid line shows the observed 
spectrum and the dashed lines show the synthetic spectra generated with $^{12}{\rm C}/^{13}{\rm C}$ equal 
to 20 (upper line) and 10 (lower line).} 
\label{fig3}
\end{figure}

The sensitivity of the abundance 
estimates to changes in the atmospheric parameters by the assumed errors is 
illustrated  for the star HD\,141680  in Table~1. 

The sensitivity of iron abundances to stellar atmospheric parameters were described 
in Puzeras et al.\ (2010).  

Since abundances of C, N and O are bound together by the molecular equilibrium 
in the stellar atmosphere, we have also investigated how an error in one of 
them typically affects the abundance determination of another. 
$\Delta{\rm [O/H]}=-0.10$ causes 
$\Delta{\rm [C/H]}=-0.04$ and $\Delta{\rm [N/H]}=0.10$;  
$\Delta{\rm [C/H]}=-0.10$ causes $\Delta{\rm [N/H]}=0.14$ and 
$\Delta{\rm [O/H]}=-0.03$. $\Delta {\rm [N/H]}=0.10$ has no effect
on either the carbon or the oxygen abundances. 

Abundances of nitrogen were determined from 14--20 lines in the spectra obtained on the Nordic Optical 
Telescope and from 9--13 lines in the spectra obtained at the Beijinh Astronomical Observatory. 
The mean scatter of the deduced line abundances is equal to 0.07~dex. This gives an approximate estimate 
of uncertainties due to random errors of the analysis.   
      
   \begin{table}
      \caption{Effects on derived abundances resulting from model changes 
for the star HD\,141680. The table entries show the effects on the 
logarithmic abundances relative to hydrogen, $\Delta$[A/H] } 
      \[
         \begin{tabular}{lrrcc}
            \hline
            \noalign{\smallskip}
Species & ${ \Delta T_{\rm eff} }\atop{ -100 {\rm~K} }$ & 
            ${ \Delta \log g }\atop{ +0.3 }$ & 
            ${ \Delta v_{\rm t} }\atop{ +0.3 {\rm km~s}^{-1}}$ &
            ${ \Delta {\rm Total}}$ \\ 
            \noalign{\smallskip}
            \hline
            \noalign{\smallskip}
   C\,(C$_2$)  & 0.02 &  0.03 &  0.00 & 0.02\\
   N\,(CN)    &--0.07 &  0.01 &  0.00 & 0.04 \\
   O\,([O{\sc i}]) & 0.01 &--0.05 &--0.01 & 0.03 \\  

   
                 \noalign{\smallskip}
            \hline
         \end{tabular}
      \]
   \end{table}


In Fig.~2 and 3, we present several examples of spectral syntheses and comparisons 
to the observed spectra.

\section{Results and discussion}

The abundances relative to hydrogen
[El/H]\footnote{In this paper we use the customary spectroscopic notation
[X/Y]$\equiv \log_{10}(N_{\rm X}/N_{\rm Y})_{\rm star} -
\log_{10}(N_{\rm X}/N_{\rm Y})_\odot$}, 
C/N,  $^{12}{\rm C}/^{13}{\rm C}$, stellar masses and suggested evolutionary stages 
determined for the programme 
stars are listed in Table~2. For convenience, we also present the main atmospheric parameters of stars 
($T_{\rm eff}$, log~$g$,  $v_{t}$ and [Fe/H]) determined by Puzeras et al.\ (2010) as well. 

\begin{table*}
\centering
\begin{minipage}{150mm}
\caption{Atmospheric parameters and chemical element abundances of the programme stars}
\begin{tabular}{rcrcrrrrrcccc}
\hline
HD & $T_{\rm eff}$ & log~$g$ & $v_{t}$ & [Fe/H] & [C/H] & [N/H] &[O/H] & C/N & $^{12}{\rm C}/^{13}{\rm C}$ & Sp. & Mass & Evol.\\
   & K             &       & km s$^{-1}$   &      &  & & & & &    & $M_{\odot}$   & \\
 \hline
    2910 & 4730 & 2.3 & 1.7 & --0.07 & --0.24 &   0.19 & --0.11 & 1.46 & 19 & 1 & 1.9 & g$^*$ \\
    3546 & 4980 & 2.0 & 1.4 & --0.60 & --0.92 & --0.30 &     -- & 0.96 & 13 & 1 & 1.5 & c \\
    4188 & 4870 & 2.9 & 1.2 &   0.10 & --0.08 &   0.23 &   0.06 & 1.95 & 10 & 2 & 2.0 & c \\
    5268 & 4870 & 1.9 & 1.4 & --0.48 &     -- &     -- & --0.33 &   -- &  --& 1 & 1.6 & c \\
    5395 & 4870 & 2.1 & 1.3 & --0.34 & --0.61 & --0.04 & --0.18 & 1.08 & 23 & 1 & 1.7 & g \\
    6805 & 4530 & 2.0 & 1.5 & --0.02 & --0.13 &   0.09 & --0.06 & 2.39 & 13 & 1 & 1.1 & c \\
    6976 & 4810 & 2.5 & 1.6 & --0.06 & --0.30 &   0.15 & --0.10 & 1.42 & 19 & 1 & 2.0 & g$^*$ \\
    7106 & 4700 & 2.4 & 1.3 &   0.02 & --0.13 &   0.35 & --0.02 & 1.32 & 19 & 1 & 1.7 & g$^*$ \\
    8207 & 4660 & 2.3 & 1.4 &   0.09 & --0.15 &   0.37 & --0.15 & 1.20 & 22 & 1 & 1.8 & g \\
    8512 & 4660 & 2.1 & 1.5 & --0.19 & --0.39 & --0.08 & --0.23 & 1.94 & 10 & 1 & 0.8 & c \\
    8763 & 4660 & 2.2 & 1.4 & --0.01 & --0.15 &   0.19 &     -- & 1.82 & 14 & 1 & 1.5 & c \\
    9408 & 4780 & 2.1 & 1.3 & --0.28 & --0.50 &   0.10 & --0.23 & 1.00 & 14 & 1 & 1.3 & c \\
   11559 & 4990 & 2.7 & 1.5 &   0.04 & --0.21 &   0.21 & --0.10 & 1.51 & 14 & 2 & 2.2 & c \\
   12583 & 4930 & 2.5 & 1.6 &   0.02 &    --  &     -- & --0.02 &   -- &  7 & 1 & 1.9 & c \\
   15779 & 4810 & 2.3 & 1.2 & --0.03 & --0.25 &   0.25 &     -- & 1.26 & 19 & 1 & 2.2 & g$^*$ \\
   16400 & 4800 & 2.4 & 1.3 &   0.00 & --0.28 &   0.35 & --0.14 & 0.93 & 20 & 1 & 2.1 & g$^*$ \\
   17361 & 4630 & 2.1 & 1.4 &   0.03 & --0.26 &   0.22 & --0.11 & 1.33 & 23 & 1 & 1.6 & g \\
   18322 & 4660 & 2.5 & 1.4 & --0.04 & --0.23 &   0.10 &   0.02 & 1.86 & 13 & 2 & 1.4 & c \\
   19787 & 4760 & 2.4 & 1.6 &   0.06 & --0.06 &   0.18 &   0.12 & 2.29 & 15 & 2 & 1.8 & c \\
   25604 & 4770 & 2.5 & 1.6 &   0.02 & --0.13 &   0.19 & --0.01 & 1.90 & 15 & 2 & 1.9 & c \\
   28292 & 4600 & 2.4 & 1.5 & --0.06 & --0.12 & --0.02 &   0.01 & 3.16 & 15 & 2 & 1.0 & c \\
   29503 & 4650 & 2.5 & 1.6 & --0.05 & --0.12 &   0.21 &   0.00 & 1.86 & -- & 2 & 1.0 & c \\
   34559 & 5060 & 3.0 & 1.5 &   0.07 & --0.13 &   0.35 &   0.03 & 1.32 & -- & 2 & 2.8 & g \\
   35369 & 4850 & 2.0 & 1.4 & --0.21 & --0.44 &   0.12 & --0.01 & 1.20 & 25 & 1 & 1.9 & g \\
   58207 & 4800 & 2.3 & 1.2 & --0.08 & --0.35 &   0.21 & --0.22 & 1.10 & 20 & 1 & 1.8 & g$^*$ \\
  131111 & 4740 & 2.5 & 1.1 & --0.17 & --0.32 &   0.05 &     -- & 1.70 & 30 & 1 & 1.3 & g \\
  141680 & 4900 & 2.5 & 1.3 & --0.07 & --0.30 &   0.21 & --0.02 & 1.24 & 16 & 1 & 2.0 & c \\
  146388 & 4700 & 2.5 & 1.3 &   0.18 &   0.03 &   0.58 &   0.04 & 1.13 & 22 & 1 & 2.0 & g \\
  203344 & 4730 & 2.4 & 1.2 & --0.06 & --0.15 &   0.17 &   0.09 & 1.92 & 22 & 1 & 1.7 & g \\
  212943 & 4660 & 2.3 & 1.2 & --0.24 & --0.41 & --0.03 &     -- & 1.66 & 30 & 1 & 1.1 & g \\
  216228 & 4740 & 2.1 & 1.3 & --0.05 & --0.30 &   0.17 & --0.19 & 1.35 & 15 & 1 & 1.7 & c \\
  218031 & 4780 & 2.3 & 1.3 & --0.08 & --0.33 &   0.07 & --0.12 & 1.59 & 13 & 1 & 1.7 & c \\
  221115 & 5000 & 2.7 & 1.3 &   0.05 & --0.26 &   0.39 & --0.09 & 0.89 & 19 & 1 & 2.5 & g$^*$ \\
  222842 & 4980 & 2.8 & 1.3 & --0.02 & --0.31 &   0.34 & --0.16 & 0.89 & 25 & 1 & 2.4 & g \\

\hline

\end{tabular}
\medskip
Sp: 1 -- NOT, 2 -- Beijing. Evol.: g -- first ascent giant, c -- He-core-burning star, $^*$ -- may be a He-core-burning star.

\end{minipage}
\end{table*}
 
\input epsf
\begin{figure*}
\epsfxsize=18,5cm 
\epsfbox[5 10 210 120]{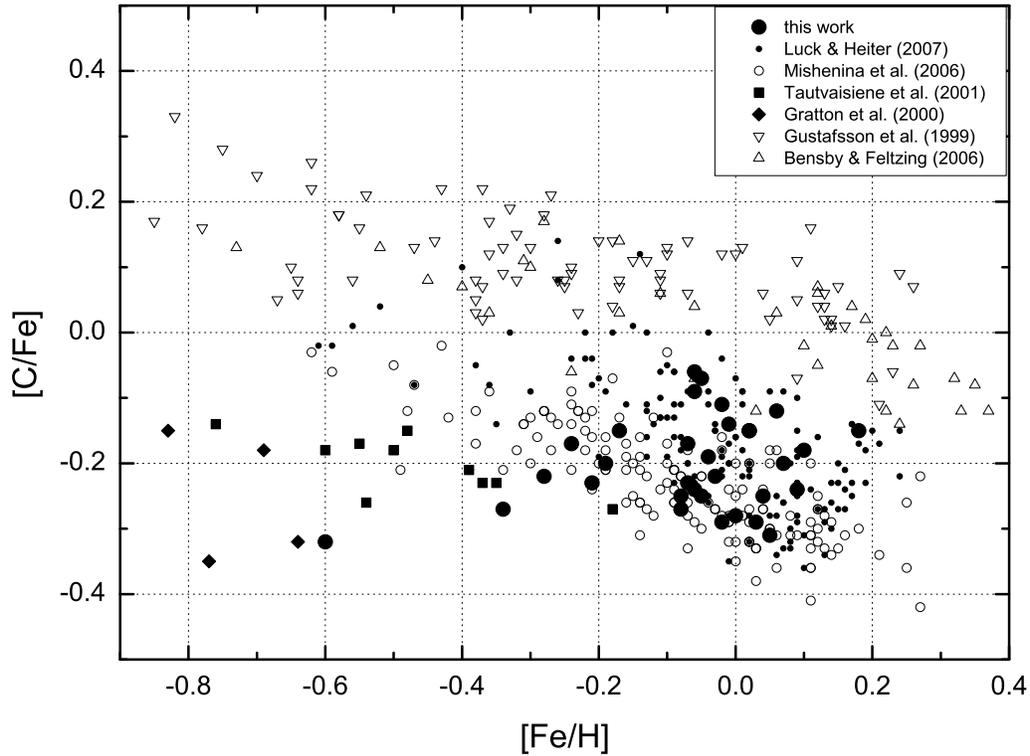} 
    \caption{[C/Fe] as a function of [Fe/H]. We show the results for 
the clump stars investigated in this work, in Mishenina et al.\ (2006) and in 
Luck \& Heiter (2007). Also we show the results obtained for red horizontal 
branch stars by Tautvai\v{s}ien\.{e} et al.\ (2001) and by Gratton et al.\ (2000). For the comparison, results obtained 
for dwarf stars of the Galactic disk (Gustafsson et al.\ 1999 and Bensby \& Feltzing 2006) are presented.}
\label{fig4}
\end{figure*}

\input epsf
\begin{figure*}
\epsfxsize=18,5cm 
\epsfbox[5 10 210 120]{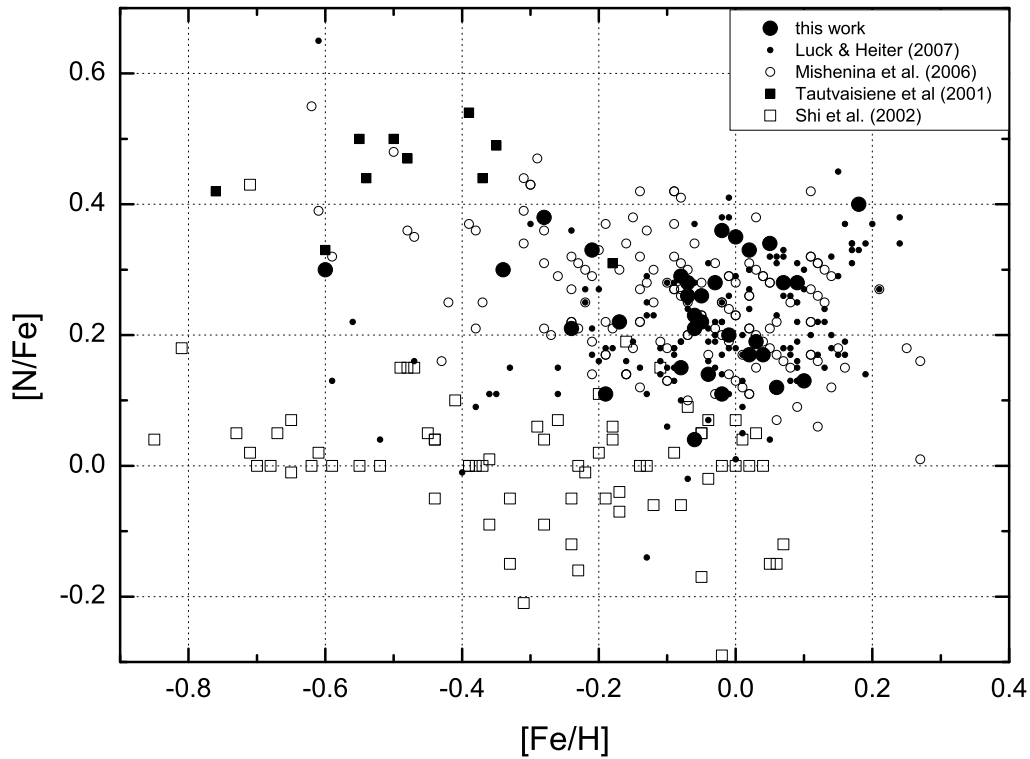} 
    \caption{[N/Fe] as a function of [Fe/H]. We show the results for 
the clump stars investigated in this work, in Mishenina et al.\ (2006) and in 
Luck \& Heiter (2007). Also we show the results obtained for red horizontal 
branch stars by Tautvai\v{s}ien\.{e} et al.\ (2001). For the comparison, results obtained for
    dwarf stars of the Galactic disk (Shi et al.\ 2002) are presented.}
\label{fig5}
\end{figure*}

\subsection{Comparisons with C, N and O abundances in dwarf stars} 

The interpretation of the C, N and O abundances can be done by a 
comparison with abundances  determined for dwarf stars 
in the Galactic disk.   

As concerns carbon, we selected for the comparison the papers by Gustafsson et al.\ (1999) and 
Bensby \& Feltzing (2006) since abundances of carbon were determined in these studies using 
the same computing programs and using the forbidden [C\,{\sc i}] line at 8727~{\AA}. 
Due to its low excitation potential, the [C\,{\sc i}] line should not be sensitive to non-LTE effects 
and to uncertainties in the adopted model atmosphere parameters, contrary to what may be expected 
for C\,{\sc i} and CH lines. The $\rm{C}_2$ line, which was investigated in our work, usually gives 
compatible results with [C\,{\sc i}]. A comprehensive discussion on this subject is given by 
Gustafsson et al.\ (1999) and Samland (1998). 

In Fig.~4, we show results of [C/Fe] for the stars investigated in our work together with results from 
other studies of clump stars performed by Mishenina et al.\ (2006) and by 
Luck \& Heiter (2007). Also we show the results obtained for red horizontal 
branch stars by Tautvai\v{s}ien\.{e} et al.\ (2001) and by Gratton et al.\ (2000). 
Compared to [C/Fe] values in dwarf stars (Gustafsson et al.\ and Bensby \& Feltzing), 
it is seen that [C/Fe] in clump stars lie by about 0.2~dex below the abundance trend of dwarfs. 

Determinations of nitrogen abundances in Galactic disk stars are not numerous. 
For metal-abundant stars, as follows from the compilation by Samland (1989), 
a concentration of [N/Fe] ratios with a rather large scatter lies at the solar value 
in the [Fe/H] interval from $+0.3$~dex to about $-1.0$~dex. 

For the comparison of [N/Fe] values, in Fig.~5, we show the results of clump stars together 
with results obtained for dwarf stars by Shi et al.\ (2002). By means of spectral synthesis 
they investigated several week N\,{\sc i} lines. Reddy et al.\ (2003) also investigated nitrogen 
abundances in a sample of 43 F--G dwarfs in the Galactic disk by means of weak N\,{\sc i} lines, 
however they used the equivalent width method, which gave, to our understanding, slightly 
overabundant [N/Fe] values. 

As it is seen from Fig.~5, in the clump stars investigated, when compared to the Galactic field dwarf 
stars, the nitrogen abundances are enhanced by about 0.2~dex.
 
In our study, as well as in Mishenina et al.\ (2006), in Liu et al.\ (2007) and 
in Luck \& Heiter (2007), the oxygen abundances in clump stars are similar to those observed in dwarfs 
(e.g. Edvardsson et al.\ 1993). 
In agreement with theoretical predictions the investigated stars do not yet show signs of evolution of 
the oxygen abundances after the main sequence. This allows to 
use oxygen abundances of clump stars for Galactic evolution studies.

\input epsf
\begin{figure*}
\epsfxsize=22cm 
\epsfbox[-55 5 800 630]{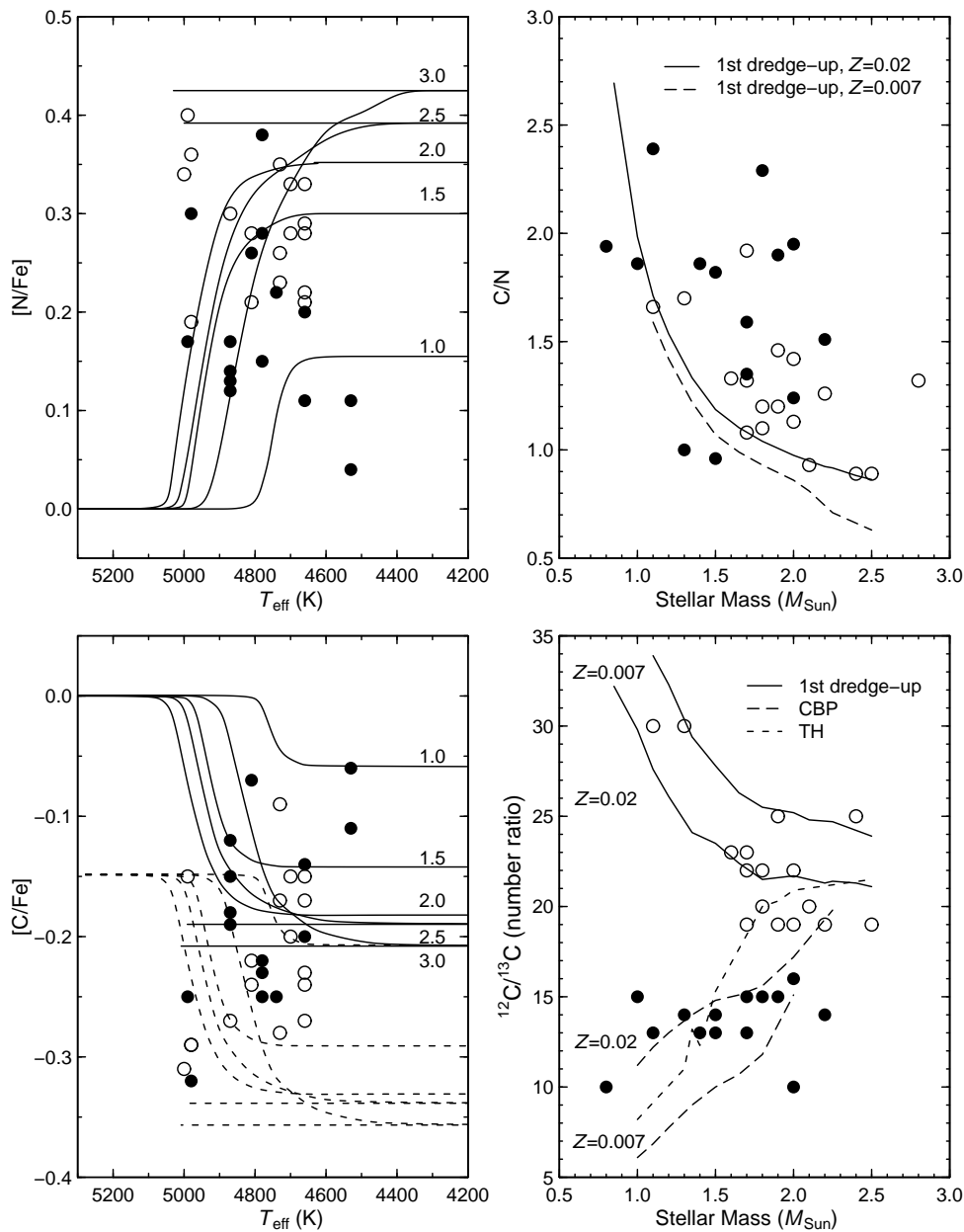} 
\caption{On the left side -- [C/Fe] and [N/Fe] versus effective temperatures in the sample of Galactic red clump stars compared 
with the theoretical tracks of abundance variations taken from Mishenina et al.\ (2006). The theoretical [C/Fe] 
(dashed lines) are shifted by $-0.15$~dex (following Mishenina et al.\ (2006) in order better to represent the 
majority of metal-deficient stars investigated. On the right side -- C/N and $^{12}{\rm C}/^{13}{\rm C}$ ratios 
versus stellar mass. The theoretical curves are taken from Boothroyd \& Sackmann (1999) -- solid and long dashed lines 
and Lagarde \& Charbonnel (2009) -- short dashed line. The stars identified as first ascent giants are shown by empty 
circles, and the stars identified as helium-core-burning stars -- filled circles.} 
\label{figx6}
\end{figure*}

\subsection{Comparisons with theoretical models}

In Mishenina et al.\ (2006), the observational results of [C/Fe] and [N/Fe] were compared with theoretical 
trends of the 1$^{st}$ dredge-up, computed using the {\sc starevol} code and presented in the same paper by 
Mishenina et al. The modelled trends were computed using the standard mixing length theory. 
In the left side of Fig.~6, we plotted [C/Fe] and [N/Fe] versus effective temperatures in our sample of 
Galactic red clump stars compared with the theoretical tracks of abundance variations taken from 
Mishenina et al. 
The nitrogen overabundances in the clump stars are in agreement with the modeled, however 
carbon in the observed sample is depleted more than the theoretical model of Mishenina et 
al.\ (2006) predicts. In these models neither overshooting, nor undershooting was considered for convection. 
The atomic diffusion and rotational-induced mixing were also not taken into account.  
In order to better fit the observational results, the authors simply shifted an initial [C/Fe] 
by $-0.15$~dex. In our comparison we had to make the same. 

C/N and $^{12}{\rm C}/^{13}{\rm C}$ ratios we compared with the theoretical models by 
Boothroyd \& Sackmann (1999) (the right side of Fig.~6) and found a good agreement with the observational data. 
These models include the deep circulation mixing below the base of 
the standard convective envelope, and the consequent "cool bottom processing" (CBP) of CNO isotopes. 

Recently Eggleton et al.\ (2006) found a mean molecular weight ($\mu$) inversion in their $1 M_{\odot}$ stellar 
evolution model, occurring after the so-called luminosity bump on the red giant branch, when the H-burning shell 
source enters the chemically homogeneous part of the envelope. The $\mu$-inversion is produced by the reaction 
$^3{\rm He(}^3{\rm He,2p)}^4{\rm He}$, as predicted by Ulrich (1972). It does not occur earlier, because the 
magnitude of the $\mu$-inversion is small and negligible compared to a stabilizing $\mu$-stratification. 
  
The work by Eggleton et al.\ (2006) has inspired Charbonnel \& Zahn (2007) to compute stellar models 
including the prescription by Ulrich (1972) and extend them to the case of a non-perfect gas for the turbulent 
diffusivity produced by that instability in stellar radiative zone. They found that a double diffusive instability 
referred to as thermohaline convection, which has been discussed long ago in the literature (Stern 1960), 
is important in evolution of red giants. This mixing connects the convective envelope with the external wing of 
hydrogen burning shell and induces surface abundance modifications in red giant stars (Lagarde \& Charbonnel 2009). 

In our $^{12}{\rm C}/^{13}{\rm C}$ and stellar mass plot (Fig.~6) we show the thermohaline model (TH) by Lagarde \& 
Charbonnel (2009) as well. It fits to our observational results well at lower masses but at larger masses the 
theoretical $^{12}{\rm C}/^{13}{\rm C}$ ratios could be slightly lower. Nevertheless, we are sure that the 
thermohaline model is a promising model to be developed. 
Cantiello \& Langer (2010) reported that thermohaline mixing is also 
present in red giants during core He-burning and beyond.           
 
The comparison with theoretical models shows that according to $^{12}{\rm C}/^{13}{\rm C}$ isotope ratios, the 
stars fall into two groups: the one with 
carbon isotope ratios altered according to the 1$^{st}$ dredge-up prediction, and the other one with carbon 
isotope ratios altered by extra mixing. The stellar positions in the 
$^{12}{\rm C}/^{13}{\rm C}$ versus stellar mass diagram as well as comparisons to stellar evolutionary 
sequences in the luminosity versus effective temperature diagram by Girardi et al.\ (2000) show that stars 
fall to groups of helium-core-burning and first ascent giants in approximately equal numbers. In the last 
column of Table~2, we indicate the predicted evolutionary status of investigated stars. By asterisks are marked 
stars which have $^{12}{\rm C}/^{13}{\rm C}$ isotope ratios equal to about 20, and which could belong to 
helium-core-burning stars as well, especially if compared to the model of thermohaline mixing.  
    
In the paper by 
Mishenina et al.\ (2006), according to nitrogen abundance values, 
the authors have suggested 21 helium-core-burning stars, about 54 candidates to helium-core-burning stars 
and about 100 first ascent giants. 

\subsection{Summary}

In this work we present $^{12}{\rm C}/^{13}{\rm C}$ determinations in a sample of 34 Galactic clump 
stars as well as abundances of nitrogen, carbon and oxygen, obtained using high-resolution spectra 
observed on the Nordic Optical telescope ($R\approx 80\,000$) and at the Beijing Astronomical Observatory
($R\approx 37\,000$). 
 
The obtained stellar abundances together with results of other studies of Galactic clump stars 
(Mishenina et al.\ 2006; Liu et al.\ 2007 and Luck \& Heiter 2007) and of red-horizontal-branch stars 
(Tautvai\v{s}ien\.{e} et al.\ 2001 and Gratton et al.\ 2000) 
were compared with determinations of C, N and O abundances in dwarf stars of the Galactic disk.  
The mean abundances in the investigated clump stars suggest that carbon is depleted by about 0.2~dex, 
nitrogen is enhanced by 0.2~dex and oxygen is close to abundances in dwarfs.  
  
The stellar positions in the 
$^{12}{\rm C}/^{13}{\rm C}$ versus stellar mass diagram as well as comparisons to stellar evolutionary 
sequences in the luminosity versus effective temperature diagram by Girardi et al.\ (2000) 
show that the stars fall into two groups: the one is of first ascent 
giants with carbon isotope ratios altered according to the 1$^{st}$ dredge-up prediction, and the other 
one is of helium-core-burning stars with carbon isotope ratios altered by extra mixing. 
The stars investigated fall to these groups in approximately equal numbers.  

And finally, we would like to point out that thermohaline convection is a fundamental physical process and 
is important in evolution of red giants. However, most probably thermohaline mixing is not the only 
physical process responsible for surface abundance anomalies in red giants (c.f. Cantiello \& Langer 2010). 
We hope that the results presented in this work will contribute to answering  
fundamental questions of stellar evolution.  

\section*{Acknowledgments}

We wish to thank Kjell Eriksson (Uppsala Observatory) for valuable help in running the synthetic spectrum 
computing programs. 
This project has been supported by the European Commission through the Baltic Grid project as well as 
through ``Access to 
Research Infrastructures Action" of the ``Improving Human Potential Programme", awarded 
to the Instituto de Astrof\' isica de Canarias to fund European Astronomers' access to the 
European Nordern Observatory, in the Canary Islands. BE thanks the Swedish Research Council (VR) for support.

\bsp

\label{lastpage}

\end{document}